\documentclass[aps,prl,letterpaper,preprint,floatfix,superscriptaddress]{revtex4-1}

\usepackage{amsmath, textcomp, graphicx, mathrsfs}
\usepackage{subfiles}

\begin{document}

\title{Direct Bandgap Light Emission from Strained Ge Nanowires Coupled with High-Q Optical Cavities}

\author{Jan Petykiewicz}
\thanks{These two authors contributed equally to this work}
\affiliation{Department of Electrical Engineering, Stanford University, Stanford, California 94305, USA}
\author{Donguk Nam}
\thanks{These two authors contributed equally to this work}
\affiliation{Department of Electronic Engineering, Inha University, Incheon 402-751, South Korea}
\author{David S. Sukhdeo}
\affiliation{Department of Electrical Engineering, Stanford University, Stanford, California 94305, USA}
\author{Shashank Gupta}
\affiliation{Department of Electrical Engineering, Stanford University, Stanford, California 94305, USA}
\author{Sonia Buckley}
\affiliation{Department of Electrical Engineering, Stanford University, Stanford, California 94305, USA}
\author{Alexander Y. Piggott}
\affiliation{Department of Electrical Engineering, Stanford University, Stanford, California 94305, USA}
\author{Jelena Vu\v{c}kovi\'{c}}
\thanks{Corresponding authors: J.V.: jela@stanford.edu, K.C.S. saraswat@stanford.edu}
\affiliation{Department of Electrical Engineering, Stanford University, Stanford, California 94305, USA}
\author{Krishna C. Saraswat}
\thanks{Corresponding authors: J.V.: jela@stanford.edu, K.C.S. saraswat@stanford.edu}
\affiliation{Department of Electrical Engineering, Stanford University, Stanford, California 94305, USA}

\begin{abstract}
A silicon-compatible light source is the final missing piece for completing high-speed, low-power on-chip optical interconnects. In this paper, we present a germanium-based light emitter that encompasses all the aspects of potential low-threshold lasers: highly strained germanium gain medium, strain-induced pseudo-heterostructure, and high-Q optical cavity. Our light emitting structure presents greatly enhanced photoluminescence into cavity modes with measured quality factors of up to 2,000. The emission wavelength is tuned over more than 400 nm with a single lithography step. We find increased optical gain in optical cavities formed with germanium under high ($>$2.3\%) tensile strain. Through quantitative analysis of gain/loss mechanisms, we find that free carrier absorption from the hole bands dominates the gain, resulting in no net gain even from highly strained, n-type doped germanium.
\end{abstract}

\maketitle
A group IV light source is the long-standing holy grail of integrated photonics, promising to enable monolithic integration of silicon (Si) CMOS electronics with high-speed and low-power optical systems \cite{Miller:867687, GeInter2014}. Attempts at Si-compatible emitters have included emission from Si nanowires \cite{Cho2013}, Si Raman lasers \cite{Takahashi2013, Rong2005}, and germanium (Ge) lasers \cite{WirthsS.2015, Camacho-Aguilera:12, Liu:10}. Of these sources, Ge offers the prospect of electrically pumped lasers, a crucial component of a fully integrated optoelectronic system.

Both optically and electrically pumped Ge lasers have been reported in the literature \cite{Liu:10, Camacho-Aguilera:12, Koerner:15}, using a combination of 0.2\% tensile strain and heavy n-type doping to achieve optical gain \cite{Liu:07}. However, the high thresholds (30 kW/cm$^2$, 280 kA/cm$^2$) and very large sizes of these lasers make them impractical for most on-chip applications. In addition, a recent report contradicts these findings \cite{PhysRevLett.109.057402}, noting that the pump-induced absorption in 0.2\% strained Ge is too high to permit lasing at the reported pump powers, indicating that it is imperative to reduce the lasing threshold. More recently, a germanium-tin (GeSn) laser was demonstrated \cite{WirthsS.2015}; however, it also suffered from similarly high threshold power (325 kW/cm$^2$) and only operated at cryogenic temperatures ($<$90 K).

Mechanical tensile strain can address the problems of both Ge and GeSn lasers by altering the bandstructure to increase the electron population in the direct valley, and thus the optical gain, resulting in greatly reduced laser thresholds \cite{Nam:11, Boucaud:13, Capellini:14, Dutt:6327582, Nam:6678709, Dave:modeling}. However, most reported strained-Ge devices exhibit only very slight strains \cite{Camacho-Aguilera:12, Prost:15, Al-Attili} and could at best provide a minimal reduction in threshold \cite{Dutt:6327582, Nam:6678709}. Attempts at inducing higher tensile strains have been made \cite{ghrib:13, SP22112011, Suess2013}, but have not included a high quality factor (Q) optical cavity and have thus merely exchanged material losses for external optical losses. Similarly, carrier confinement in a double-heterostructure has been explored as an avenue for achieving lasing, but requires complex material growth and is often incompatible with tensile strain \cite{Sun:10, Chen:2014}.

In this work, we propose a novel structure that addresses all three issues jointly: high tensile strain for improved material gain, a compact and high-Q optical nanocavity, and a pseudo-heterostructure. We present a new nanocavity design capable of confining both light and excited carriers in a $>$2.3\% tensile-strained active region while maintaining quality factors up to 2000. Our structures require only a single lithography step and enable tunable emission over a wavelength range of more than 400 nm. By combining our photoluminescence data with simulations of optical losses in our devices, we extract the net optical gain induced by our pump laser. We find increased net optical gain in Ge nanowires with higher tensile strain, and we examine the loss mechanisms which inhibit lasing in our highly strained devices. We fit our results to band-filling calculations and extract a minority carrier lifetime of 2.9 ns, in good agreement with prior time-resolved photoluminescence (PL) measurements \cite{Nam:14}.

\begin{figure}[htbp]
\centering
\includegraphics[width=16cm]{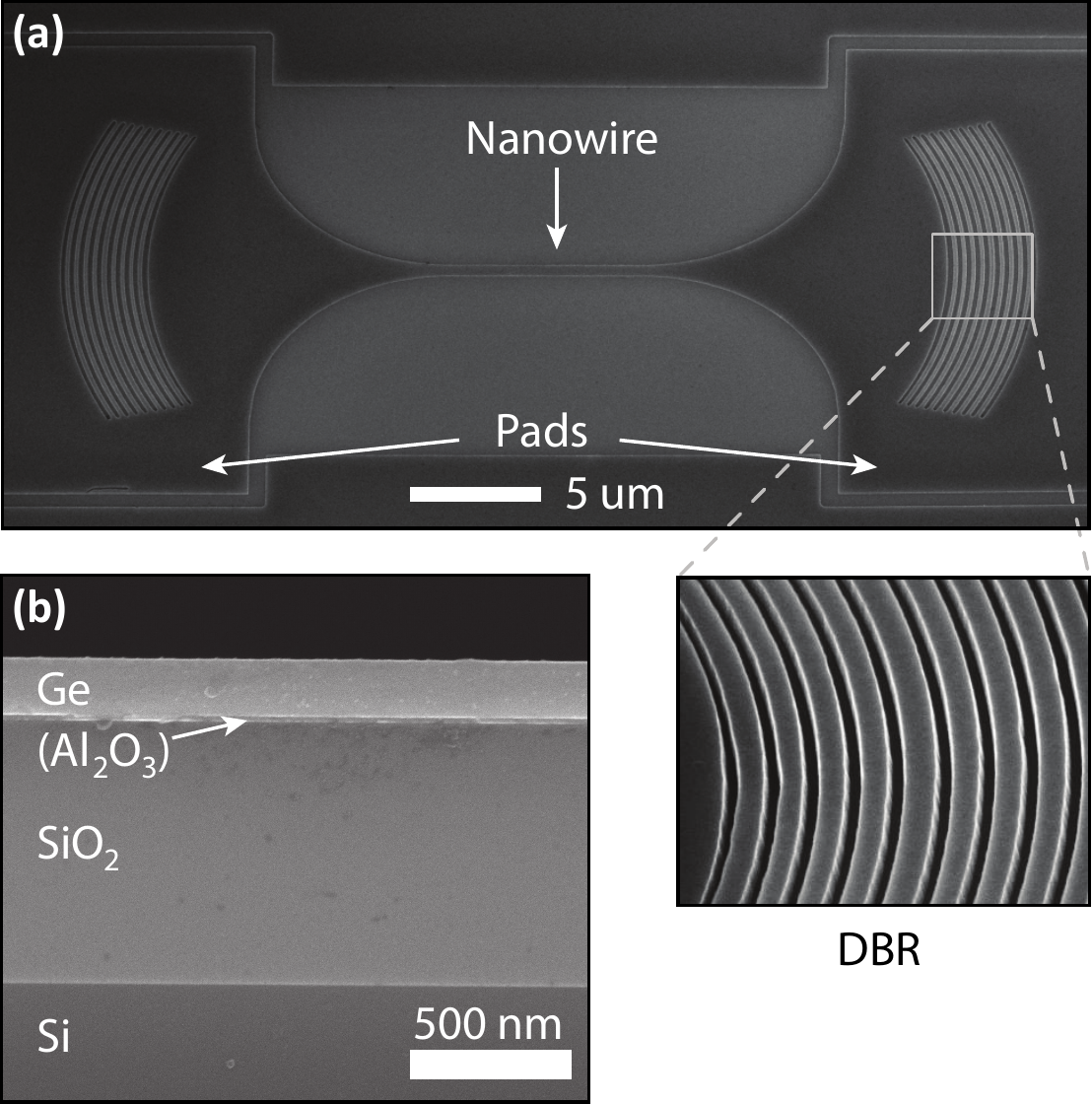}
\caption{Strained Ge nanowire. (a) Scanning electron micrograph (SEM) of a fabricated device, showing the etched Ge nanowire and adjacent distributed Bragg-reflector (DBR) mirrors (detail view). (b) Side-view SEM of the initial Ge-on-insulator material stack.}
\label{fig1}
\end{figure}

Figure 1a presents a scanning electron micrograph (SEM) of a fabricated device, consisting of an 8 \textmu m long highly-strained Ge nanowire surrounded by two large pads containing distributed Bragg reflector (DBR) mirrors which form the optical cavity. The device is fabricated in a material stack consisting of four layers: 200 nm Ge, 25 nm Al$_2$O$_3$, 850 nm SiO$_2$, and Si wafer (figure 1b). Fabrication details for creating the material stack are provided in the supplemental material (SM). The Ge layer is n-type doped with $1\times 10^{19}$ cm$^{-3}$ phosphorous atoms using in-situ doping during growth. The device is patterned using a single electron beam lithography step, followed by a HBr/Cl$_2$ dry etch for pattern transfer to the Ge layer. The structure is then undercut by a KOH wet etch to selectively remove the Al$_2$O$_3$ layer, releasing the very slightly ($0.2\%$) strained Ge layer from the substrate. This step allows the large pads to contract, amplifying the strain in the nanowire \cite{Suess2013, Nam:PSH, Sukhdeo:14}.

While crucial for inducing strain in the nanowire, the undercut creates an air gap between the Ge layer and the substrate, severely limiting thermal conduction out from the nanowire and resulting in the destruction of devices under optical pump powers of $\sim$5 mW. To address this limitation, we make use of capillary forces to pull the Ge nanowire down and bring it into contact with the SiO$_2$ layer when drying the device after the undercut step. The nanowire is then held in contact with the SiO$_2$ layer by van der Waals forces. The SiO$_2$ layer enables optical confinement in the Ge while also serving as an additional heat conduction path, permitting a $>$10x increase in pump power. Thermal conduction simulation results are included in the SM (figure S2). Figures 2a,b present height-maps of several devices, gathered using white-light interferometry. Figure 2c presents a line-scan across the single device indicated in figure 2b. A depression of ~25 nm can be seen in the vicinity of each device, equal to the thickness of the sacrificial Al$_2$O$_3$ layer. Finally, the reattached Ge nanowire is coated with a thin ($\sim$10 nm) conformal Al$_2$O$_3$ layer using atomic layer deposition (ALD), passivating the surface and further improving thermal characteristics. Figure 2d presents a side-view schematic of the final device.

\begin{figure}[htbp]
\centering
\includegraphics[width=16cm]{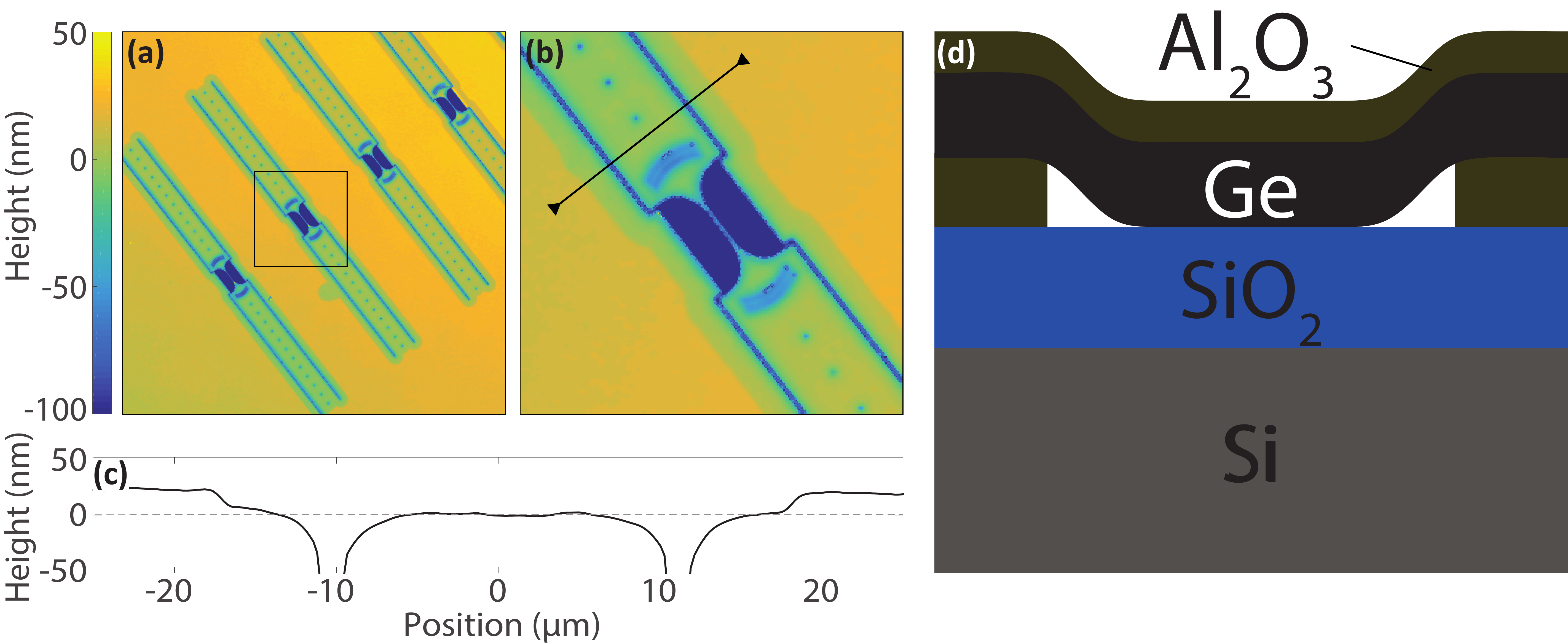}
\caption{Substrate stiction. (a) Wide-area and (b) detailed height-maps of the wafer surface, gathered using white-light interferometry. The boxed area from (a) is shown in (b). The devices and nearby undercut regions are lower than the wafer surface. Height values below -100 nm are clamped for clarity. (c) Line-scan of the region in (b),  showing a step height of $\sim$25 nm at the edge of the undercut region, corresponding to the thickness of the sacrificial Al$_2$O$_3$ layer. (d) Side-view schematic of the final material stack. Layer thicknesses are 20 nm Al$_2$O$_3$, 200 nm Ge, 25 nm Al$_2$O$_3$, 850 nm SiO$_2$, Si substrate.}
\label{fig2}
\end{figure}

Using a combination of finite difference time domain (FDTD) optical simulations (figure 3a) and finite element method (FEM) mechanical modeling (figure 3b), we designed a device that supports optical modes with radiative quality factors of over $10^4$ while retaining very high ($>$2\%) mechanical strain along the nanowire. The straight 8 \textmu m long x 700 nm wide active region is expanded to a maximum width of 13.7 \textmu m and then connected to 20 \textmu m-wide side pads. The total distance between mirrors is $\sim$30 \textmu m. 10-period DBR mirrors with a period of 380 nm and a nominal duty cycle of 21\% are matched to the shape of the optical mode. We provide additional details about the mirror design in the SM.

\begin{figure}[htbp]
\centering
\includegraphics[width=16cm]{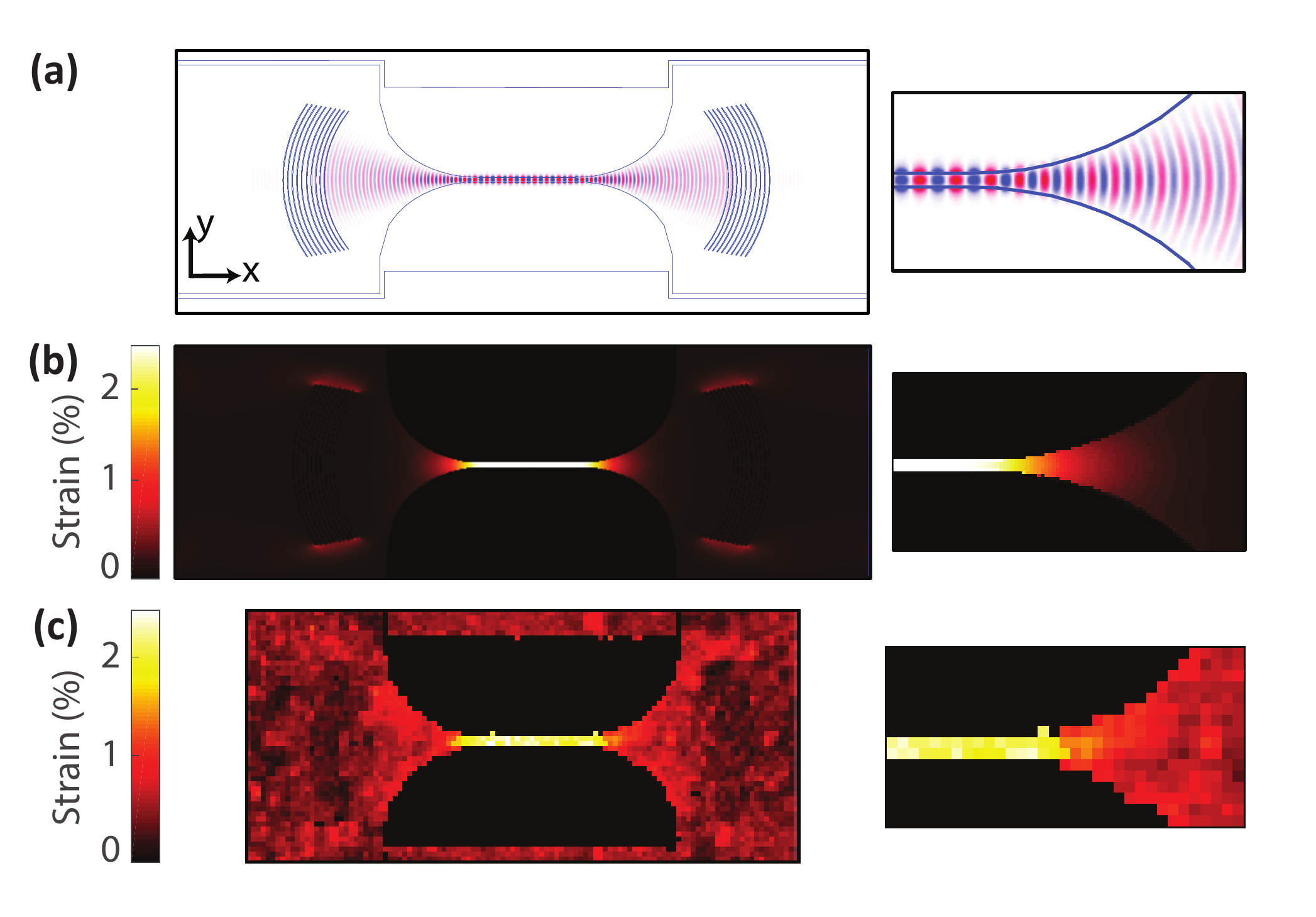}
\caption{Optical mode and strain distribution. Insets on right show a magnified view of the right side of the wire and adjacent tapered region. (a) Ey field distribution for the optical mode with a free-space wavelength of 2.0 \textmu m, obtained with FDTD simulation. (b) Tensile strain distribution in an identical structure, obtained with FEM simulation. A uniform strain of $\sim$2.4\% is present along the narrow central segment of the nanowire. (c) Strain map for a fabricated device, experimentally obtained using Raman spectroscopy.}
\label{fig3}
\end{figure}

From simulations, we find that the strain is very uniform along the straight central region of the nanowire as well as in the out-of-plane direction throughout the device. We calculate an optical confinement factor of 0.37 in the maximally strained region for the mode pictured in figure 3a. Figure 3c shows an experimentally measured strain distribution in a fabricated device, collected using Raman spectroscopy (details in SM). The result matches our FEM simulations and reveals uniform highly strained Ge along the length of the nanowire. Additionally, because tensile strain reduces the bandgap of Ge, the spatial variation in the strain profile creates a pseudo-hetereostructure which confines carriers to the nanowire region, greatly improving carrier concentrations in the gain medium \cite{Nam:PSH}.

\begin{figure}[htbp]
\centering
\includegraphics[width=16cm]{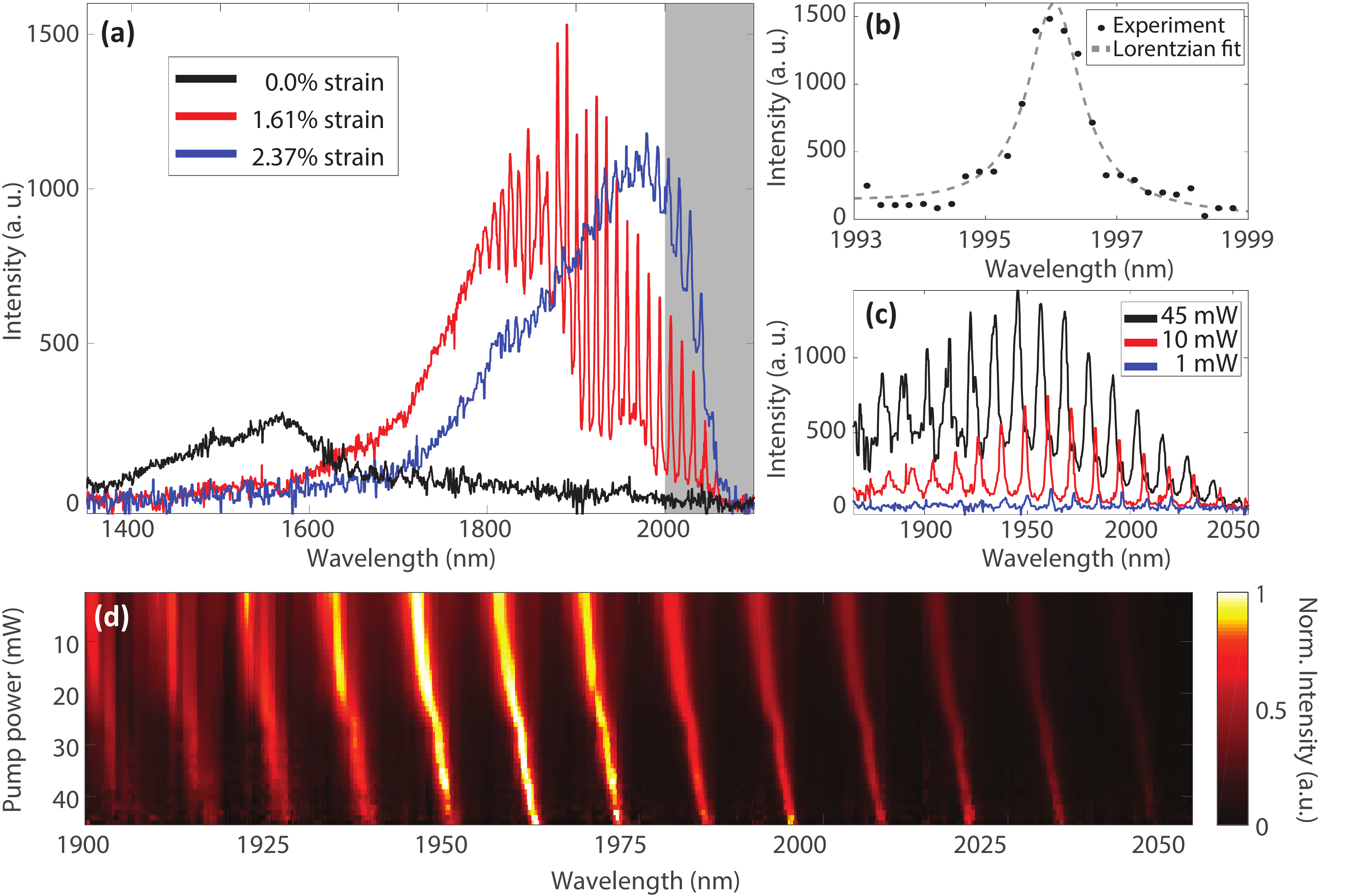}
\caption{Photoluminescence (PL). (a) PL spectra for devices with 0\%, 1.61\%, and 2.37\% tensile strain. The direct-gap emission intensity increases greatly at higher strain and exhibits high-Q resonances with a free spectral range of 24.6 nm. The detection limit of the spectrometer is denoted with a gray background. (b) Lorentzian fit to a single resonance peak with fitted quality factor of 2020. (c) High resolution PL spectra for a 1.95\% strained nanowire pumped with increasing average power from a pulsed laser source. Results at wavelengths below $\sim$1950 nm include atmospheric absorption lines. (d) Spectral dependence of the emission on the pump power. Each spectrum (row) is normalized to its peak value. A large free-carrier-induced blue shift and significant broadening are observed for all modes.}
\label{fig4}
\end{figure}

Figure 4a shows photoluminescence spectra from unstrained, 1.61\%, and 2.37\% strained nanowires with identical optical cavities. The devices are pumped with 2.5 mW of power from a continuous-wave (CW) 980 nm diode laser. PL intensities at wavelengths longer than 2000 nm are reduced due to the detection limit of the extended InGaAs detector array used to collect the spectra. The emission peak can be seen to redshift from $\sim$1560 nm to at least 1950 nm as the strain in the nanowire increases and the bandgap narrows \cite{Nam:6678709}. Optical resonances with a free spectral range of 24.6 nm are visible in the strained wire spectra, but are obscured by noise in the unstrained wire spectrum due to low signal intensity. Figure 4b presents a lorentzian fit to a single optical mode under identical pump conditions. A Q-factor of 2000 is observed, limited by sidewall roughness along the nanowire and material absorption.

Figure 4c presents a 1.95\% strained nanowire under pulsed excitation with various average powers, pumped by a 1550 nm laser with 2 \textmu s pulse period and 200 ns pulse length, chosen to minimize heating. Figure 4d displays a series of spectra taken with varying pump powers, with the intensity in each row normalized to its maximum value. Resonances at all wavelengths are seen to shift to shorter wavelengths and broaden as the pump power is increased; both effects are readily explained by increased free carrier densities in the active region. Contrary to previous reports \cite{Camacho-Aguilera:12, Liu:10}, we did not observe signs of lasing, such as linewidth narrowing, although our material is highly strained and has a similar doping level to the reported optically pumped Ge-on-Si laser.

To understand the gain/loss mechanisms in our highly strained Ge resonator, we quantitatively calculate the change in net material gain necessary to cause the Q reduction we observe experimentally. We express the net loss $\alpha = \frac{\omega \sqrt{2}}{c} \sqrt{\sqrt{1+(\frac{\sigma_{eff}}{\epsilon \omega})^2}-1}$, in terms of the material dielectric constant $\epsilon$, the angular frequency of the emitted light $\omega$, and an effective conductivity $\sigma_{eff}$ \cite{JacksonBook}. We then relate $\sigma_{eff}$ to the pump-dependent quality factor $Q_{abs}=\omega E / P_{abs}$, where $E$ is the energy in the resonator and $P_{abs}$ is the instantaneous absorbed power. $P_{abs}$ is expressed as $P_{abs}=\iiint\limits_{wire} \sigma_{eff} |\mathscr{E}|^2 \,dV$, where $\mathscr{E}$ is the electric field distribution; self-consistent values of $\mathscr{E}$ and $E$ were taken from FDTD simulation. Finally, we find $Q_{abs}$ by decomposing the experimentally measured quality factor $Q_{exp}$ into two components, a constant cold-cavity quality factor $Q_{cold}$ and a power-dependent component $Q_{abs}$, with $Q_{exp}^{-1} = Q_{cold}^{-1} + Q_{abs}^{-1}$ and assuming $Q_{abs} \propto P_{pump}^{-1}$ (SM figure S4).

\begin{figure}[htbp]
\centering
\includegraphics[width=16cm]{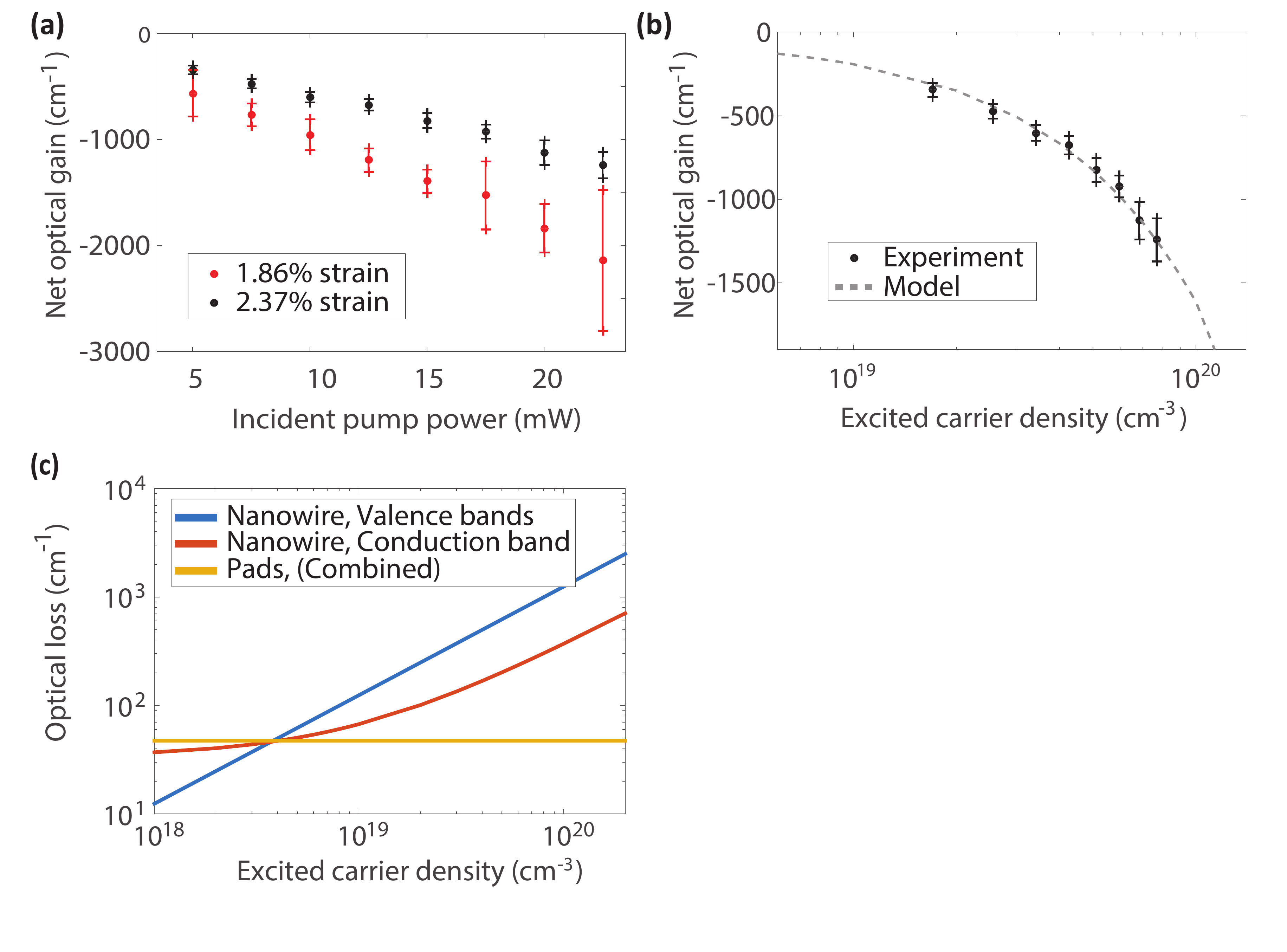}
\caption{Analysis of net optical gain. (a) Plot of net optical gain versus incident pump power (980 nm CW) for 2.37\% (black) and 1.86\% (red) strained nanowires. A clear improvement in net gain is seen in the more strained device. (b) Net optical gain versus injected carrier density for the 2.37\% strained device. Black dots indicate experimental data, and the dashed grey line indicates values calculated from band-filling simulations. (c) Plot of optical losses from free-carrier transitions for the curve presented in (b). Intra-valence-band transitions in the highly strained nanowire dominate. At high pump powers, the contribution from pad regions surrounding the nanowire is negligible due to strong carrier confinement in the pseudo-heterostructure.}
\label{fig5}
\end{figure}

We plot the net optical gain for 2.37\% and 1.86\% strained nanowires in figure 5a. The net gain is seen to decrease with increasing pump power, indicating that increasing free-carrier losses overwhelm any gross gain from the direct-band transition. Nevertheless, we find a significant improvement in net gain as the strain is increased from 1.86\% to 2.37\%, likely due to an increase in the available gross gain. By fitting joint-density-of-state (JDOS) calculations (figure 5b, dashed line) to our calculated net gain, we extract a minority carrier lifetime of 2.9 ns, in good agreement with previous measurements \cite{Nam:14}.

To clarify the dominant loss mechanisms, we present them individually in Figure 5c, as determined with our JDOS calculations. From this plot, we attribute the Q reduction we observe in our structure primarily to transitions within the hole bands. This is in line with previous observations in highly-doped Ge microdisks \cite{Shambat:10} and is supported by the pump-induced absorption reported by Carroll et al. \cite{PhysRevLett.109.057402}, though it is contradictory to existing reports of lasing in Ge \cite{Liu:10, Camacho-Aguilera:12}.

In summary, we have demonstrated greatly enhanced direct-bandgap light emission from a structure which lays the groundwork for a practical group IV laser by integrating high tensile strain, a pseudo-heterostructure, and a low-loss optical cavity. We have presented resonances with Q factors of up to 2,000 and emission tunable over a $>$400 nm range in devices fabricated with a single lithography step. We have performed a quantitative analysis of the pump-dependent net optical gain in our resonators and found increased net optical gain from our highly strained devices. We have determined that the strain levels achieved in our devices, hole-band free carrier losses dominate any optical gain in our material. We have found good agreement between measured optical gain and band-filling calculations and extracted a minority carrier lifetime of 2.9 ns. We expect that application of further mechanical strain \cite{ Dutt:6327582, Nam:6678709} or translation of our work to a GeSn material system \cite{Dave:modeling} will enable practical Si-compatible lasers with order-of-magnitude improvements in threshold over the state of the art.

\begin{acknowledgments}
We gratefully acknowledge financial support from the AFOSR MURI on Robust and Complex On-Chip Nanophotonics (Dr. Gernot Pomrenke, Grant No. A9550-09-1-0704) and from APIC Corporation (Dr. Raj Dutt). This research was also supported by the Pioneer Research Center Program through the National Research Foundation of Korea funded by the Ministry of Science, ICT \& Future Planning (2014M3C1A3052580). J. P. was suported in part by the National Physical Science Consortium Fellowship with stipend support from the National Security Agency. A.Y.P. acknowledges support from the Stanford Graduate Fellowship.
\end{acknowledgments}

%

\pagebreak

\widetext
\begin{center}
\textbf{\large Supplemental Materials: Direct Bandgap Light Emission from Strained Ge Nanowires Coupled with High-Q Optical Cavities}
\end{center}
\setcounter{equation}{0}
\setcounter{figure}{0}
\setcounter{table}{0}
\setcounter{page}{1}
\makeatletter
\renewcommand{\theequation}{S\arabic{equation}}
\renewcommand{\thefigure}{S\arabic{figure}}
\renewcommand{\bibnumfmt}[1]{[S#1]}
\renewcommand{\citenumfont}[1]{S#1}

\section{Material stack fabrication}
The fabrication process flow for the initial material stack is depicted in figure S1. First, a 2.2 \textmu m-thick Ge layer with an n-type dopant concentration of $1\times 10^{19}$ cm$ ^{-3}$ was grown on a Si substrate using multiple hydrogen annealing heteroepitaxy (MHAH) \cite{tmah}. A 25 nm thick Al$_2$O$_3$ sacrificial layer was then deposited using atomic layer deposition (ALD). A Si carrier wafer with an 850 nm thick SiO$_2$ layer was prepared by thermal oxidation at 1100 \textdegree C. After wet-chemical treatment of the two bonding surfaces (Al$_2$O$_3$ and SiO$_2$), the two wafers were directly bonded at room temperature, followed by a high-temperature anneal to increase bonding strength. The Si substrate of the carrier wafer was then selectively etched away using TMAH, stopping at the Ge layer. By using chemical-mechanical polishing (CMP), the Ge layer was thinned down to 200 nm, finalizing the fabrication process for the substrate.

\begin{figure}[htbp]
\centering
\includegraphics[width=16cm]{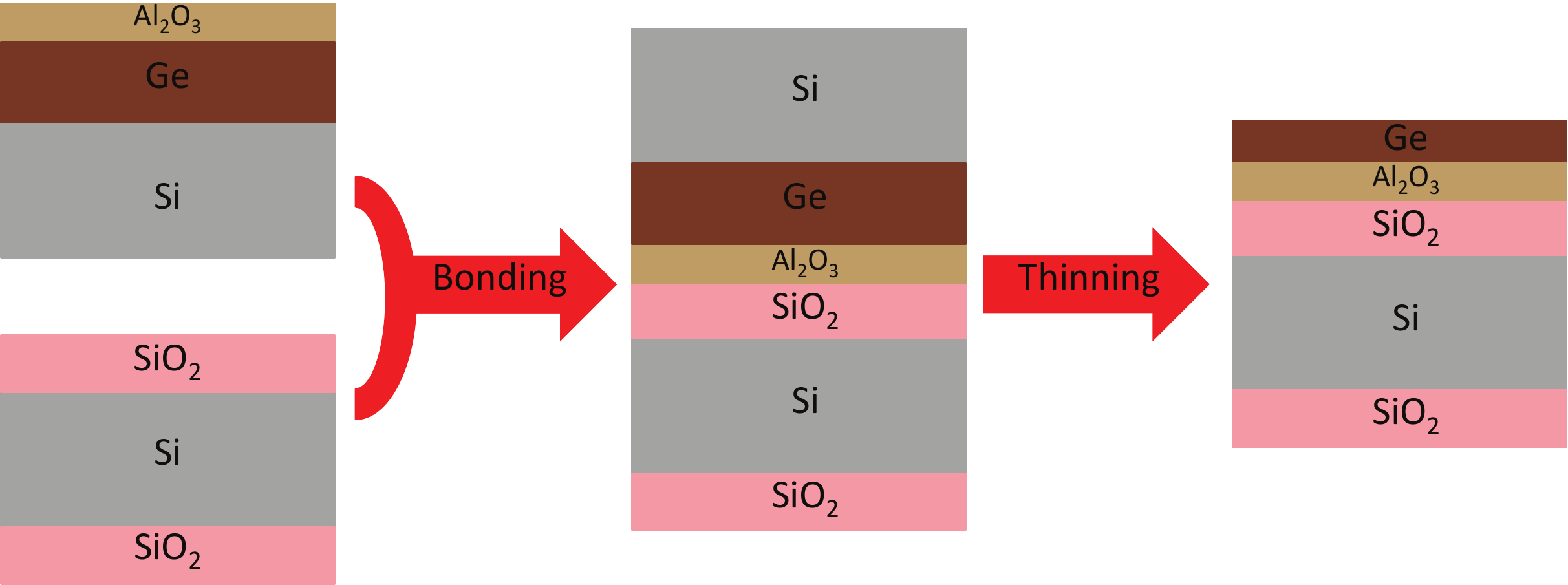}
\caption{Substrate fabrication process. The left-most panel shows the initial growth of Ge and deposition of Al$_2$O$_3$ on Si, as well as the growth of SiO$_2$ on a separate Si wafer. These wafers are then bonded, the resulting top Si layer removed, and the Ge layer thinned to create the material stack used to begin device fabrication.}
\label{sfigs1}
\end{figure}

\section{Thermal simulations}
Figure S2 presents the thermal advantages of bringing the undercut Ge device layer into contact with the underlying SiO$_2$ layer rather than leaving it suspended in air. Figure S2a presents an FEM simulation of an air-suspended 200 nm Ge membrane containing a 10 mW, 1 \textmu m-radius circular heat source, 20 \textmu s after the heat source is switched on. Meanwhile, figure S2b presents the same membrane, in contact with the remainder of the material stack (SiO$_2$ and Si substrate). The suspended membrane reaches temperatures over 1200 K, with a maximum of 1575 K at the spot center. Meanwhile, the majority of the adhered substrate remains relatively cool ($\sim$375 K; maximum 526 K). Figure S2c displays the time evolution of the temperature at the center of the heat source location, showing a $>3\times$ improvement in steady-state temperature as well as suggesting a \textmu s-order timescale for any potential heating effects.

\begin{figure}[htbp]
\centering
\includegraphics[width=10cm]{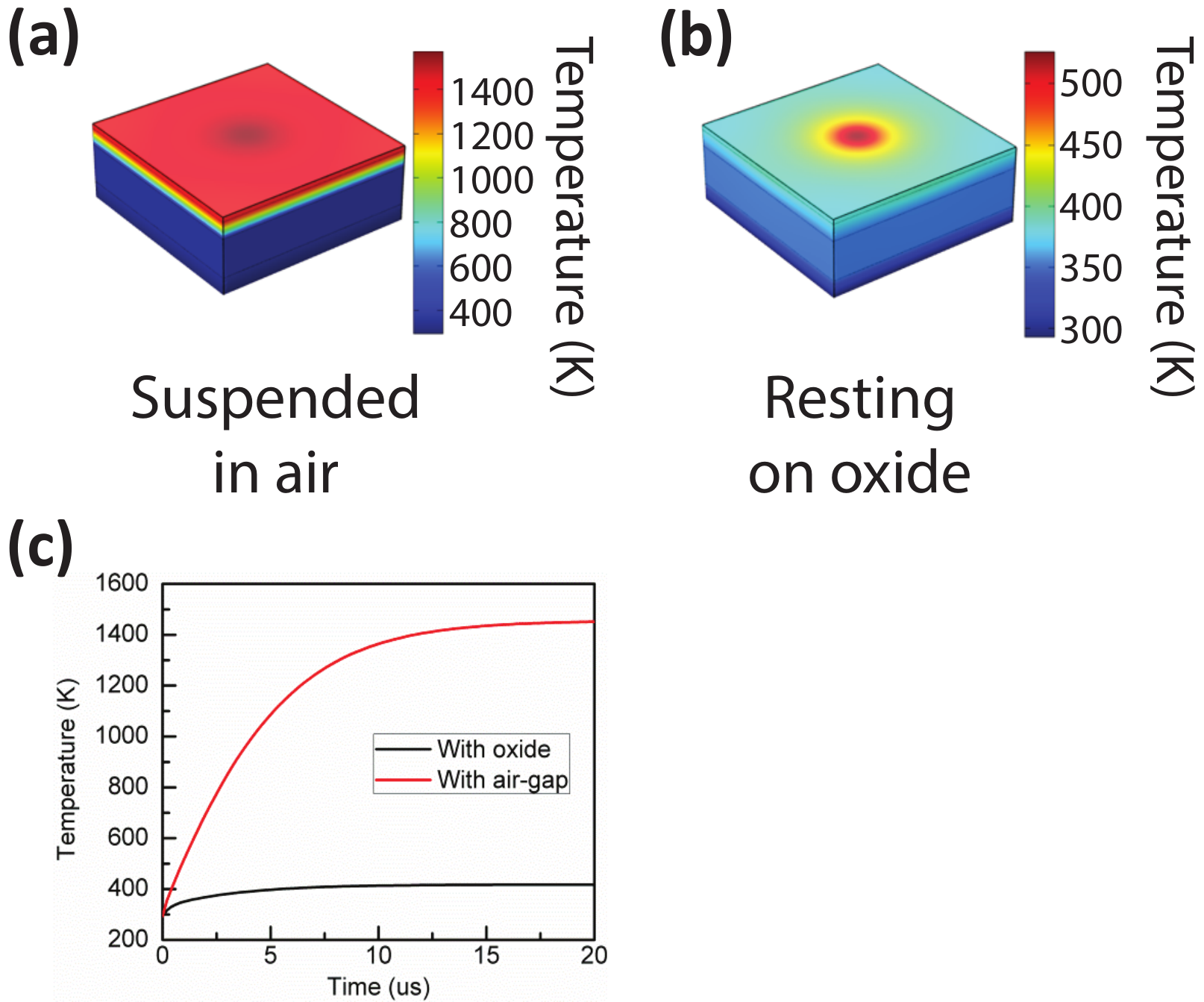}
\caption{Thermal simulations. Temperature maps for a 200 nm Ge membrane (a) suspended in air, and (b) in contact with the underlying SiO$_2$ layer. Plots are taken after 20 \textmu s of elapsed time, during which a 1 \textmu m-radius circular heat source generates 10 mW at near the top center of the plotted region. (c) Time evolution of the temperature of the Ge membrane near the center of the circular heat source. The source is switched on at $t = 0$.}
\label{thermal}
\end{figure}

\section{Raman spectroscopy}
Raman spectroscopy was performed by focusing a 514 nm laser source onto the sample using a 100x objective lens. The excitation power was kept low ($<$100 \textmu W) during the measurement to ensure no significant heating from the laser. The sample stage was scanned in two dimensions, and the measured Raman shift at each position was converted to an equivalent uniaxial strain using a strain-shift coefficient of 152 cm$^{-1}$ \cite{Suess2013}.

\section{DBR mirror design}
Each DBR mirror in our optical cavity consists of 10 etched trenches with a nominal periodicity of 380 nm and duty cycle of 21\%, resulting in a nominal trench width of 80 nm. The trenches follow circular arcs, with a nominal radius of curvature equal to $0.63d$, where $d$ is the distance from the end of the straight portion of the nanowire to the center-point on the arc. The outermost arc is positioned at $d = 7.89$ \textmu m. Finally, the first three trenches are perturbed to further increase the simulated Q factor:
\begin{enumerate}
\item Moved outward by $\delta d = $ 30 nm and narrowed to a width of 60 nm.
\item Moved outward by $\delta d = $ 8 nm and narrowed to a width of 69 nm.
\item Moved inward by $-\delta d = $ 25 nm.
\end{enumerate}
Additional mirror periods can be added to further suppress in-plane radiative losses, but this was not necessary in this work.

\section{Cavity Q calculations}
\begin{figure}[htbp]
\centering
\includegraphics[width=10cm]{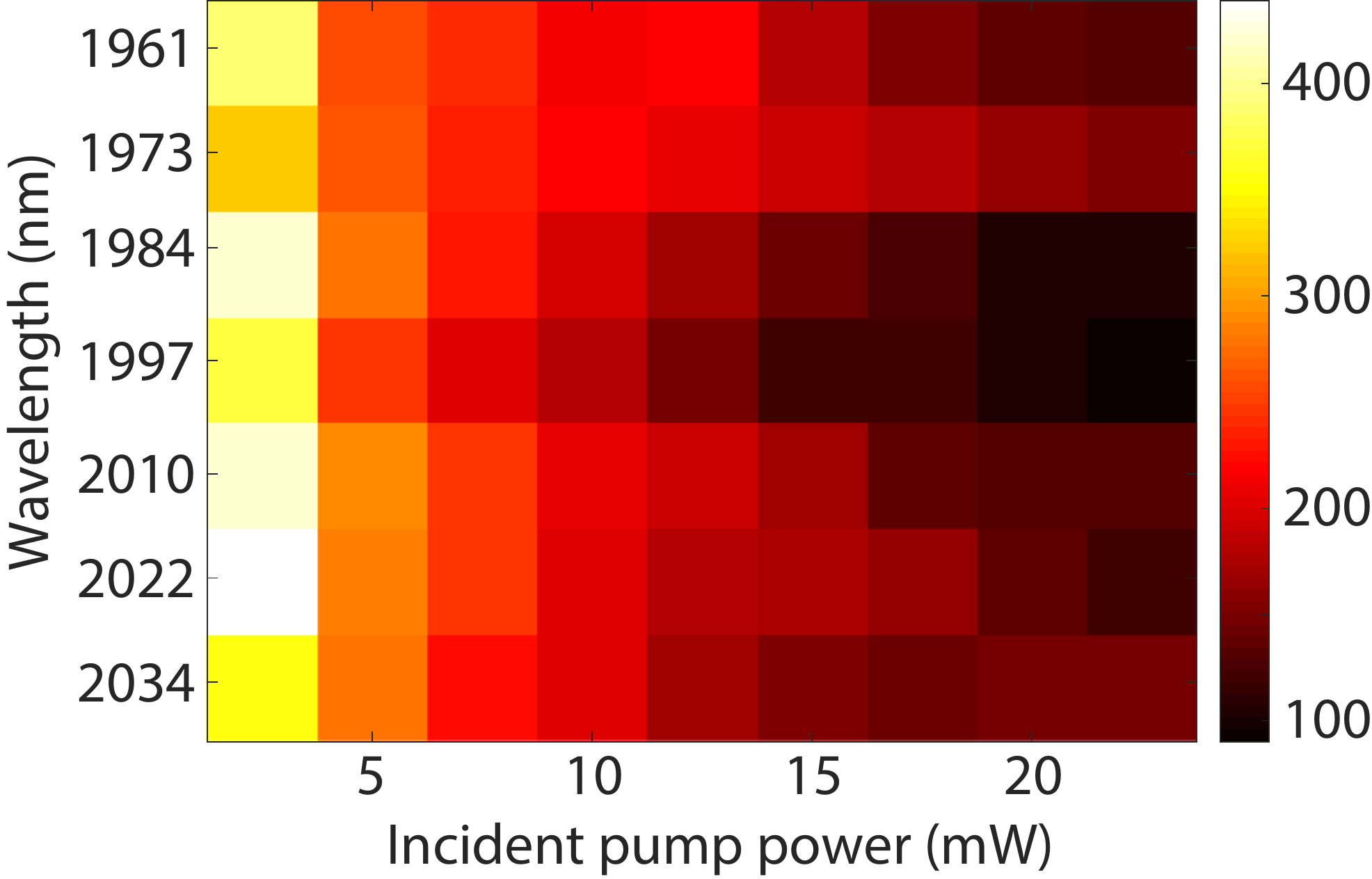}
\caption{Quality factors for multiple modes for the 2.37\%-strained device at various incident pump powers. Due to the high strain of the device, only relatively low-Q modes appear in the detection window (1950-2040 nm).}
\label{qmap}
\end{figure}

\begin{figure}[htbp]
\centering
\includegraphics[width=10cm]{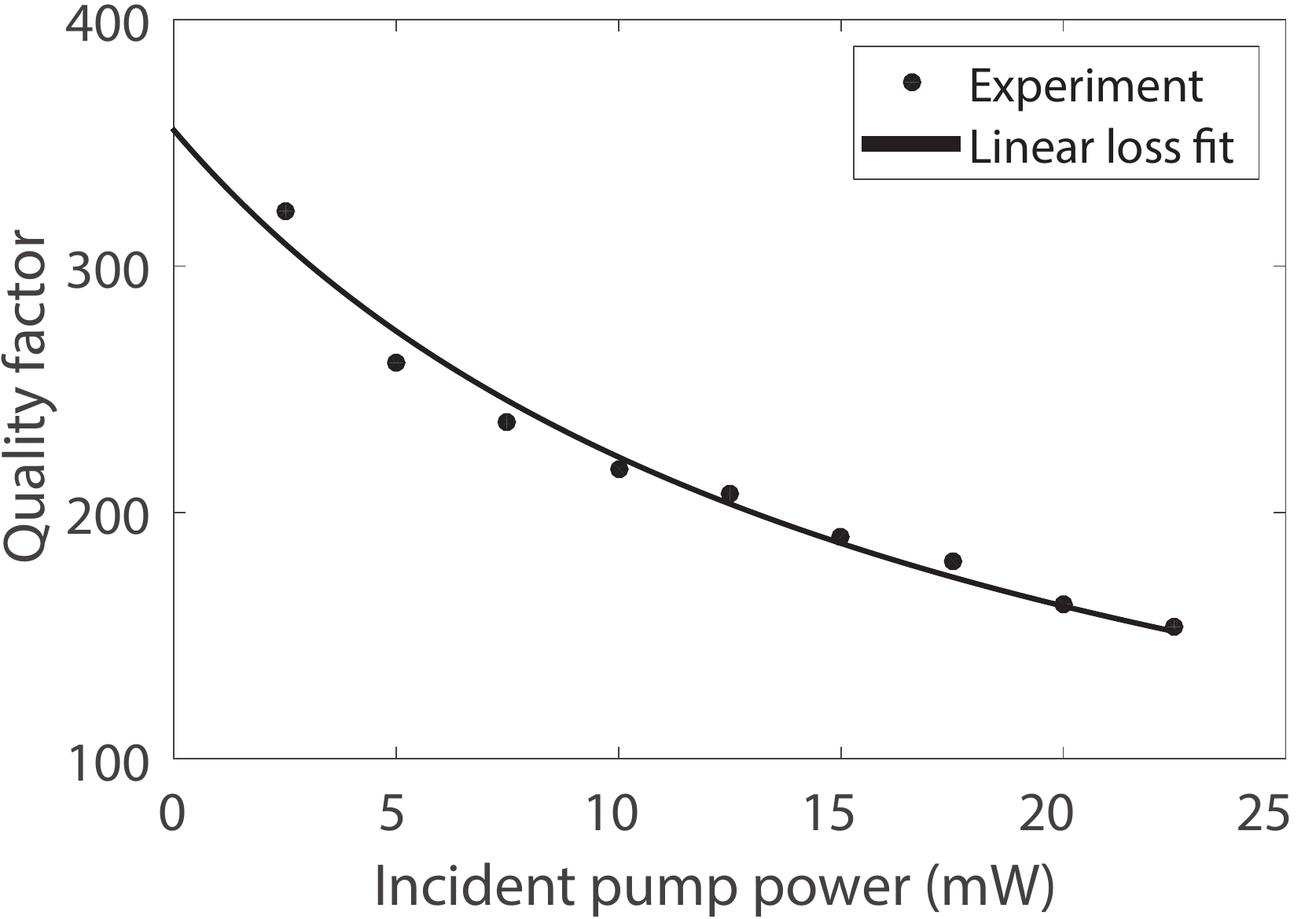}
\caption{Example least-squares fit to cavity quality factor versus incident pump power. Dots correspond to measured quality factor for the device with 2.37\% strain, (i.e. a row in figure S3). The solid line corresponds to the fitted curve, as discussed in the main text.}
\label{sfigs2}
\end{figure}

Figure S3 presents the measured quality factor $Q_{exp}$ for the 2.37\% strained device. To determine the cold-cavity Q factor, a least-squares fit with linear power-dependent loss is performed, as pictured in figure S4. The fitted curve is of the form $Q_{exp}^{-1} = Q_{cold}^{-1} + Q_{abs}^{-1}$, where $Q_{abs} \propto P_{pump}^{-1}$, as discussed in the main text.

\pagebreak

\end{document}